# Parameter Estimation of Gaussian-Damped Sinusoids from a Geometric Perspective

Thomas A. Pelaia II - ORCID: 0000-0002-5879-9340
*Oak Ridge National Lab, Oak Ridge, TN 37831, USA*

## ABSTRACT

The five parameter gaussian damped sinusoid equation is a reasonable model for betatron motion with chromatic decoherence of the proton bunch centroid signal in the ring at the Spallation Neutron Source. A geometric method for efficiently fitting this equation to the turn by turn signals to extract the betatron tune and damping constant will be presented. This method separates the parameters into global and local parameters and allows the use of vector arithmetic to eliminate the local parameters from the parameter search space. Furthermore, this method is easily generalized to reduce the parameter search space for a larger class of problems.

## I. INTRODUCTION

The five parameter gaussian damped sinusoid equation is a reasonable model for the transverse motion of a single proton bunch injected into the accumulator ring at the Spallation Neutron Source (SNS) [1] at Oak Ridge National Lab and stored for several dozen turns. Each beam position monitor (BPM) in the ring can capture this turn by turn signal for each transverse plane resulting in a waveform (one per plane) with one element per turn. A waveform depends on both local beam parameters (orbit distortion, amplitude and phase) at the BPM and global beam parameters (betatron tune and damping constant) independent of BPM. Our goal is to measure the global beam parameters, and more specifically we are most interested in efficiently and accurately measuring the betatron tune with concurrently captured waveforms across all BPMs.

In this paper we present a method based on a geometric perspective to fit the global parameters to one or more waveforms eliminating the local parameters from the parameter search space. With just two global parameters to fit, the optimization complexity is significantly reduced and especially ideal when fitting to multiple waveforms across all BPMs. By eliminating the unknown local parameters from the parameter search space, it is expected the fits should

This manuscript has been authored by UT-Battelle, LLC, under Contract No. DE-AC05-00OR22725 with the U.S. Department of Energy. The United States Government retains and the publisher, by accepting the article for publication, acknowledges that the United States Government retains a non-exclusive, paid-up, irrevocable, world-wide license to publish or reproduce the published form of this manuscript, or allow others to do so, for United States Government purposes.

consistently converge to the optimal global parameters faster. It will be shown that this technique naturally extends to a wider class of problems.

## II. BETATRON MOTION WITH CHROMATIC DECOHERENCE

A charged particle at the nominal energy in the ring will exhibit simple betatron motion in the absence of nonlinearities. As is well known, if a particle has an energy that differs from the nominal, the tune will shift accordingly as the product of the relative energy shift and the chromaticity of the ring optics. For a bunch with a distribution of energies about the nominal energy, the motion of its constituent particles will decohere over time due to the corresponding tune spread. The equation of motion has previously been derived [2] for general synchrotron motion by Meller et. al; however, for our case a simpler model can be used. For times much smaller than the synchrotron period, it is straight forward to show that the bunch centroid signal (such as measured by a BPM) will be a gaussian-damped sinusoid. At SNS, this is indeed the region of relevance as the bunch signal is typically analyzed for several dozen stored turns, and the synchrotron period is greater than 1400 turns [3]. That the observed signal damping for nominal chromaticity is due to chromatic decoherence is consistent with measurements showing that the damping time increases by several dozen turns as the chromaticity is reduced to zero.

By definition, the betatron tune is shifted from the nominal tune by the product of the chromaticity and the relative energy shift. The orbit is also distorted by an amount equal to the product of the dispersion by the relative energy shift. For times much shorter than the synchrotron period, the relative energy shift is effectively constant, so we can approximate the position of a single particle as a function of turn by equation 1 where $b$ is the orbit distortion, $A$ is the amplitude, $\varphi$ is the phase, $\mu$ is $2\pi$ times the tune, $\mu_0$ is $2\pi$ times the nominal tune, $t$ is the turn index, $\eta$ is the dispersion and $\xi$ is the chromaticity.

$$q = b + A \sin(\mu t + \varphi) + \frac{\eta}{2\pi\xi}(\mu - \mu_0) \tag{1}$$

Given a bunch of particles, equation 2 shows the resulting centroid position of the bunch as a function of time in terms of the distribution function over $\mu$.

$$<q> = b + \int_{-\infty}^{\infty} d\mu \rho(\mu) \left( A \sin(\mu t + \varphi) + \frac{\eta}{2\pi\xi}(\mu - \mu_0) \right) \tag{2}$$

If the energy spread follows a gaussian distribution, then the distribution over $\mu$ can be written according to equation 3 where $\sigma_\mu$ is the standard deviation of $\mu$.

$$\rho(\mu) = \frac{1}{\sigma_\mu \sqrt{2\pi}} e^{-(\mu-\mu_0)^2 / 2\sigma_\mu^2} \tag{3}$$

Assuming this distribution and performing the integral, the centroid of motion is indeed determined to be in the form of a gaussian damped sinusoid as shown by equation 4.

$$<q> = b + Ae^{-(\sigma_\mu t)^2/2} \sin(\mu_0 t + \varphi) \qquad (4)$$

Indeed, BPM waveforms seem to be consistent with the gaussian-damped sinusoid over the first several dozen turns. Figure 1 shows a representative ring BPM waveform of a single injected bunch at nominal chromaticity which fits well to a gaussian-damped sinusoid.

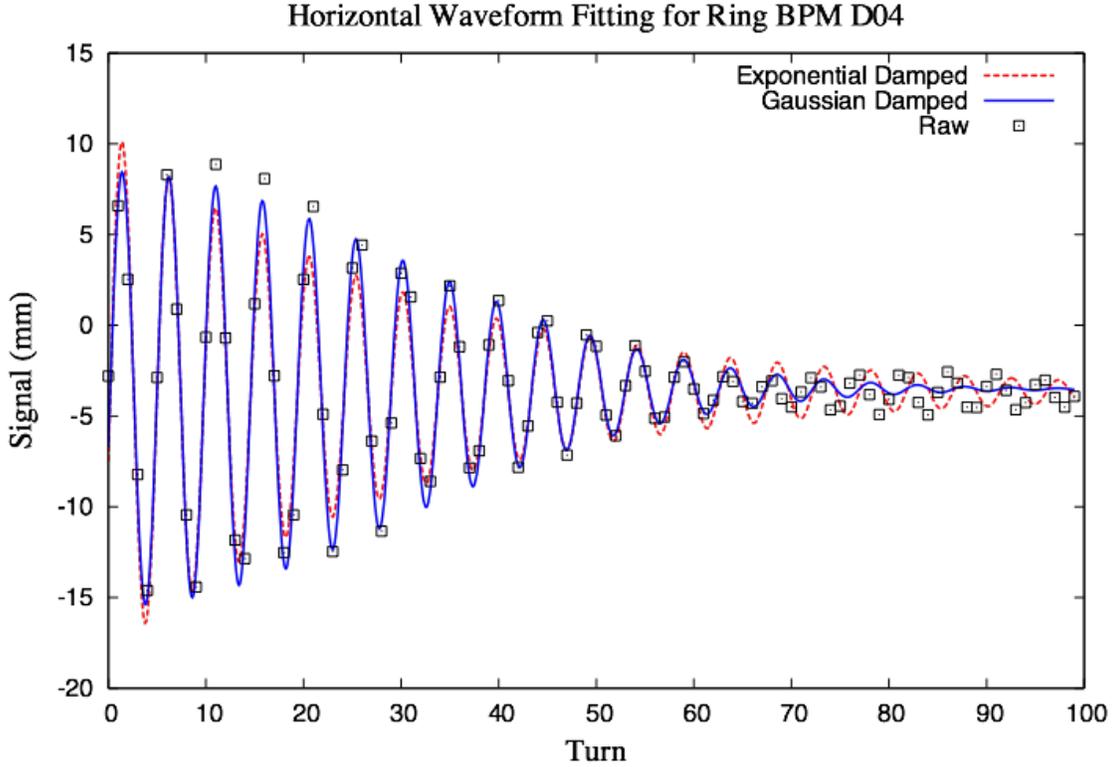

Figure 1. Representative SNS ring BPM horizontal waveform with both exponentially damped sinusoid and gaussian-damped sinusoid fits. The gaussian-damped sinusoid fit is best in the first sixty-five turns after which it damps away into the noise and other physics appears to dominate.

## III. GAUSSIAN-DAMPED SINUSOID PARAMETER ESTIMATION

Equation 5 shows the general form of the gaussian-damped sinusoid with amplitude $A$, damping constant $\gamma$, angular frequency $\mu$, phase $\varphi$ and orbit distortion (offset) $b$. The waveform signal $q_n$ is evaluated at the zero based turn index $n$.

$$q_n = Ae^{-\gamma n^2} \sin(\mu n + \varphi) + b \qquad (5)$$

By inspection there are five parameters to fit and at least that many turns within the waveform will be needed to fit the data. Two of the parameters (damping constant and frequency) are global parameters which are shared in common throughout the ring and the remaining three (amplitude, phase and orbit distortion) are local parameters which vary throughout the ring. Since we are most interested in using the BPM turn by turn signals to measure tune, we are only interested in estimating the global parameters, and this is especially true when fitting over many concurrent waveforms corresponding to the BPMs distributed throughout the ring.

Conventional parameter estimation methods based on least squares fitting to a signal typically require a search over all five parameters. Even worse, if multiple waveforms are to be fit as we wish, the search becomes unnecessarily expensive as there will be two global parameters plus three local parameters per waveform to find. The search typically also requires good initial estimates of the parameters, and good phase and amplitude estimates can be challenging to compute. However, a duality exists between least squares estimation and geometric shortest distance problems [4] that leads to the desired algorithm. The geometric viewpoint most naturally provides the mechanics for effectively eliminating the local parameters and reducing the parameter search space to just that of the global parameters.

To help visualize the geometric approach, consider the simpler problem presented in figure 2 in which three turn waveforms are fitted to simple sinusoids with (local) unknown amplitudes and phases and a common (global) unknown frequency to determine. In this case, for a specified frequency, the set of all possible amplitudes and phases form a plane passing through the origin, and the best frequency is the one for which its plane is closest to the measured waveforms represented by points in the three dimensional space.

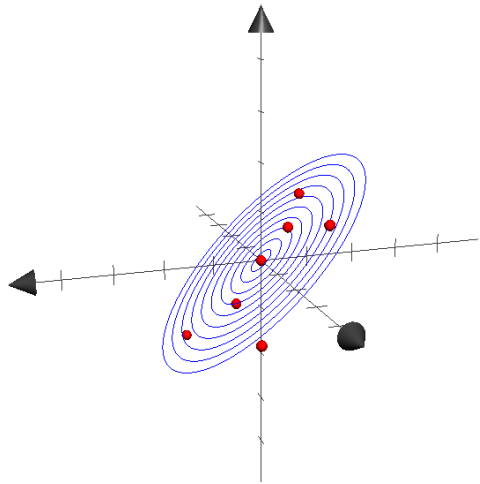

Figure 2. Three element (turn) waveforms form a three dimensional space with each measured waveform represented by a point (red dot in figure). Consider the simple model in which the waveforms are represented by sinusoids sharing a common (global) unknown frequency, but each with (local) unknown amplitude and phase. For fixed frequency, the set of amplitudes and phases form a plane passing through the origin. Varying phase forms

ellipses and varying amplitude changes the size of these ellipses all lying on a plane. The best fitting frequency is the one for which the corresponding plane passes closest to the measured waveforms (red dots).

For the five parameter gaussian-damped sinusoid problem, geometric manipulation can reduce the parameter search space to just the two global parameters (frequency and damping constant). In the geometric perspective, any waveform can be mapped one to one to a single point in a space of dimension equal to the length of the waveform where each element corresponds to a coordinate in the space. Within this space, all perfectly gaussian-damped sinusoid waveforms form a five dimensional subspace (one degree of freedom per parameter). If the two global parameters ($\mu$ and $\gamma$) are fixed, there is a smaller corresponding subspace that is three dimensional and covers all possible values of the local parameters (amplitude, phase and offset). This three dimensional subspace is in fact a three dimensional vector subspace which allows the local parameters to be eliminated from the optimization. To see that this subspace is in fact a vector subspace, we can use the simple trigonometric identity where a sine with a phase can be expressed as the sum of a sine and cosine with the same frequency and appropriate coefficients. So the gaussian-damped sinusoid can be rewritten as follows where the amplitude and phase are absorbed into the coefficients on the sine and cosine terms and the offset, $b$, is simply renamed to $r_3$ for consistency in the new form.

$$q_n = e^{-\gamma n^2}(r_1 \sin \mu n + r_2 \cos \mu n) + r_3 \qquad (6)$$

In this form, this equation can be rewritten as a vector equation with coefficients on three constant (for fixed $\mu$ and $\gamma$) basis vectors where the vector elements correspond to the turns indexed by $n$.

$$\vec{q} = r_1 \vec{S}_{\mu,\gamma} + r_2 \vec{C}_{\mu,\gamma} + r_3 \vec{Z} \qquad (7)$$

$\vec{S}_{\mu,\gamma}$ and $\vec{C}_{\mu,\gamma}$ are vectors with sine like and cosine like gaussian-damped elements that depend only on $\mu$, $\gamma$ and the turn index $n$ and $\vec{Z}$ is a constant vector of just ones. For $\mu$ strictly in the open interval from zero to $\pi$, these three basis vectors are linearly independent. Linear independence can be verified through inspection as follows. The first element of $\vec{S}_{\mu,\gamma}$ is identically zero and that of $\vec{C}_{\mu,\gamma}$ and $\vec{Z}$ are identically one so there can be no nonzero factor from $\vec{S}_{\mu,\gamma}$ to $\vec{C}_{\mu,\gamma}$ or $\vec{Z}$. Since $\vec{Z}$ is a vector of all ones, there can be no common factor from $\vec{C}_{\mu,\gamma}$ to $\vec{Z}$. Also, $\vec{S}_{\mu,\gamma}$ and $\vec{C}_{\mu,\gamma}$ cannot be linearly combined to make $\vec{Z}$ because summing sine like and cosine like vectors results in another sinusoid vector with a phase shift and there can be no common factor from this resulting sinusoid vector to a constant vector. Gaussian damping cannot compensate the sinusoid terms to make them constant because it provides monotonic damping over the elements.

It is expected that a measured BPM turn by turn waveform will only be approximately gaussian-damped sinusoid due to noise and other contributions such as nonlinear effects and coupling not accounted for by our model. Such a waveform will be outside of the five dimensional solution space (and hence the three dimension $\mu$-$\gamma$ subspace for the optimal global parameters). However, such a waveform should be near the optimal µ-ɣ subspace, and so we identify the best fit by finding the two global parameters (µ and ɣ) that minimize the shortest distance from the waveform to the µ-ɣ subspace for these global parameters. We seek to develop an algorithm to directly compute this distance from a waveform to a µ-ɣ subspace without searching. Such an algorithm would effectively reduce the parameter search space from five down to just the two global parameters. It would allow for efficient global parameter estimation over several concurrent waveforms sharing the same global parameters (e.g. BPM turn by turn waveforms throughout the accumulator ring) by simply computing the root mean square (RMS) of distances over the waveforms to the µ-ɣ subspace and reducing the overall parameter search space by three local parameters per waveform. It will be shown that this measure of error is both convenient using simple vector arithmetic and identical to the usual RMS signal error validating it as the ideal measure for goodness of fit.

Geometrically, the shortest distance from a measured waveform vector, $\vec{q}_w$ (with waveform index *w*), to the µ-ɣ subspace is the magnitude of the component of $\vec{q}_w$ orthogonal to the µ-ɣ subspace. The first step is to compute $\vec{S}_{\mu,\gamma}$ and $\vec{C}_{\mu,\gamma}$ for µ and ɣ recalling that $\vec{Z}$ is just a constant vector of ones. We have already established that $\vec{S}_{\mu,\gamma}$ and $\vec{C}_{\mu,\gamma}$ and $\vec{Z}$ are linearly independent, but they are not generally orthogonal or normalized. The second step is to form an orthonormal bases, $\vec{U}_i$ (i=1,2,3), from these vectors. Orthogonalization and normalization can be done using standard procedures such as the Modified Gram-Schmidt algorithm involving basic vector arithmetic. The shortest distance is given by the magnitude of the waveform component orthogonal to these bases vectors, $|\vec{E}_w|$ where the error vector is given by the following equation.

$$\vec{E}_w = \vec{q}_w - \sum_{j=1}^{3} \vec{U}_j (\vec{q}_w \cdot \vec{U}_j) \tag{8}$$

By construction, it is clear that $\vec{E}_w$ is orthogonal to each of the basis vectors and hence to the µ-ɣ subspace and is the difference between the measured waveform vector and its projection onto this subspace. Thus it must be shortest distance from the measured waveform point to the µ-ɣ subspace. By inspection, it is also clear that $\vec{E}_w$ is identical to the signal error and thus minimizing the magnitude of $\vec{E}_w$ over µ and ɣ is equivalent to minimizing the RMS signal error over all five parameters. We can apply this method to multiple concurrent waveforms using the same bases for each µ and ɣ pair and compute the RMS distance over the waveforms as the appropriate error to minimize. Thus we have achieved our goal of reducing the parameter search space to just the global parameters and eliminating the search over the local parameters we don't need.

# IV. PERFORMANCE SIMULATIONS

Simulations were performed to both verify the algorithm and measure its performance using parameters relevant to the operation of SNS. The simulations were performed using the Open XAL [5] solver which is a black box optimizer.

The plot in Figure 3 shows the performance of the geometric parameter estimation comparing single waveform fits and multiple waveform fits based on 40 sources (since SNS has just over 40 BPMs in the ring). Using multiple concurrent waveforms to the fit the tune offers a clear advantage over fitting to a single waveform as expected. Based on statistical noise, one would expect roughly a reduction in error by a factor of the square root of the number of waveforms used in the fitting.

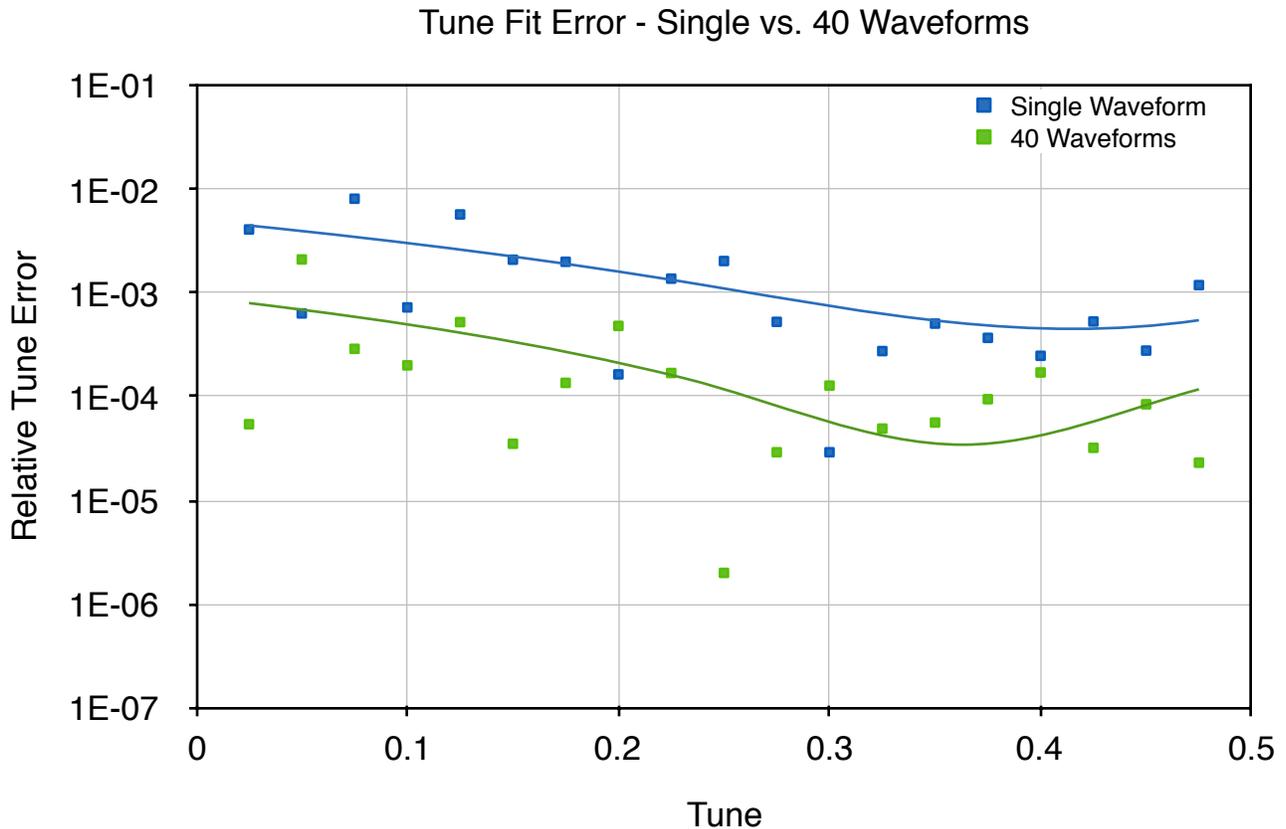

Figure 3. The log plot shows the relative tune error versus tune for both multiple waveform fits and single waveform fits. The waveforms were generated using a gaussian damping constant of 0.0005, random phases from 0 to $2\pi$, random amplitudes from 5 mm to 15 mm, random offsets ranging over ±10 mm and gaussian signal noise with standard deviation of 1 mm. The solid lines are just trend lines.

Figure 4 shows the performance of the geometric fitting algorithm compared with conventional direct least squares parameter estimation. 40 waveforms were used in the fits. In the conventional method, each waveform was fit to the gaussian damped sinusoid using the black box solver to find the five parameters that best fit to the waveform, and then the tune was determined by averaging over the tunes from all the waveforms. The total optimization time was limited to 1 second wall time for both methods per run with one run per trial tune. The geometric method consistently outperformed the conventional least squares method by roughly two orders of magnitude in the relative error for the same total amount of wall clock time.

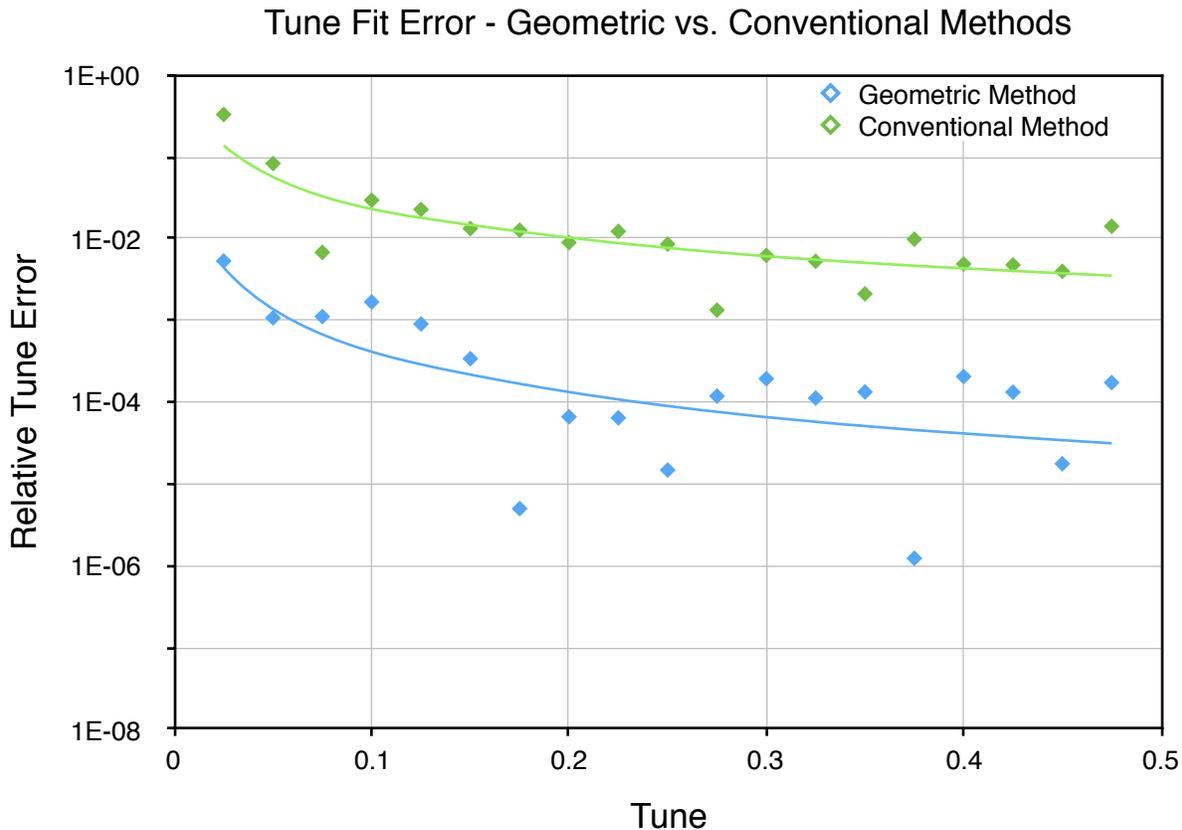

Figure 4. The log plot shows the relative tune error versus tune for the geometric and the conventional methods. For each point, the two methods were each given 1 second total wall clock optimization time using the black box solver. The waveforms were generated using a gaussian damping constant of 0.0005, random phases from 0 to $2\pi$, random amplitudes from 5 mm to 15 mm, random offsets ranging over ±10 mm and gaussian signal noise with standard deviation of 1 mm. The solid lines are just trend lines.

# V. GENERALIZATION OF GEOMETRIC PARAMETER ESTIMATION

The algorithm just presented for reducing the parameter search space using geometric parameter estimation can be generalized to fitting equations in a broader form beyond the gaussian-damped sinusoid. Consider $K$ sources of waveforms $\vec{q}_k$ of length $N$ which can be modeled as the linear sum of $J$ linearly independent vector functions $\vec{F}_j$ of $M$ unknown global parameters $\lambda_m$ and with coefficients $r_{k,j}$ which are unknown parameters local to the sources.

$$\vec{q}_k = \sum_{j=1}^{J} r_{kj} \vec{F}_j(\lambda_1, \ldots, \lambda_M) \tag{9}$$

The goal is to find the global parameters which best fit this equation to the real waveform data from all the sources. The geometric perspective allows this problem to be reduced to a search for just the global parameters. For each set of trial global parameters there exists a subspace covering all possible values for the local parameters, and the measure of fit error is taken to be the shortest distance from the measured waveforms to this subspace. Because the local parameters appear just as coefficients on linearly independent vector functions, this subspace is a vector subspace which allows the shortest distance to be computed directly with vector arithmetic.

The optimization procedure is a search over the space of global parameters, $\lambda_m$. For each set of trial global parameters, the vector functions $\vec{F}_j$ are to be computed. From these evaluated vector functions, an orthonormal bases is formed using vector arithmetic (e.g. using Modified Gram-Schmidt). The fit error for a specific waveform fit is computed to be the magnitude of the difference between the waveform vector and the projection onto the orthonormal bases.

$$E_k = \left\| \vec{q}_k - \sum_{j=1}^{J} \vec{U}_j (\vec{q}_k \cdot \vec{U}_j) \right\| \tag{10}$$

When fitting multiple concurrent waveforms with the same global parameters, compute the overall fit error as the RMS of the individual waveform fit errors. Vary the global parameters $\lambda_m$ to minimize the overall fit error. If signal noise variance among the waveform sources is known, then variance weighting can be applied as usual when computing the overall fit error.

This solution is equivalent to standard matrix based least squares estimation, but the geometric perspective provides a geometrically intuitive procedure involving vector operations to eliminate the unknown coefficients without the need for explicit matrix inversion.

## V. CONCLUSIONS

Ring BPM turn by turn waveforms can be modeled as five parameter gaussian-damped sinusoids for charged particle betatron motion with chromatic decoherence over turns much less than the synchrotron period. Two of these parameters are global (independent of BPM) and three are local (BPM dependent). A geometric perspective in which a waveform is viewed as a vector in a space of dimension equal to the length of the waveform provides important insight into efficiently solving this problem. Because the equation can be written in a form where the local parameters only appear as linear coefficients, vector arithmetic was used to eliminate the local parameters and reduce the parameter search space to be just the two global parameters.

The geometric approach was then generalized for parameter reduction of a class of problems in which the parameters to eliminate appear only as linear coefficients on linearly independent vector functions of other parameters to fit.

# REFERENCES

[1] M.A. Plum, "Commissioning Of The Spallation Neutron Source Accelerator Systems," Proceedings of PAC07, Albuquerque, New Mexico, 2007.

[2] R.E. Meller et. al., "Decoherence of Kicked Beams," SSC-N-360, 1987.

[3] The SNS Synchrotron period estimate was provided by internal communication with Michael Plum.

[4] Gilbert Strang, "Introduction to Applied Mathematics," Wellesley-Cambridge Press, 1986, pp. 34-39, 385-389.

[5] T Pelaia II et al., Open XAL Status Report 2015, http://accelconf.web.cern.ch/AccelConf/IPAC2015/papers/mopwi050.pdf, Proceedings of IPAC 2015, Richmond, VA (2015)